\documentclass[aps,prb,twocolumn,showpacs,amsmath,amssymb]{revtex4}
\usepackage{epsfig}
\usepackage{color}
\usepackage[english]{babel}
\DeclareGraphicsExtensions{.pdf,.png,.gif,.jpg,.eps}
\usepackage{bm}

\newcommand{\hlS}[1]{{\color{black}{#1}}} 
\newcommand{\hlA}[1]{{\color{black}{#1}}}   

\begin{document}
\title{Dynamical symmetry breaking in vibration-assisted transport through nanostructures}
\author{Abdullah Yar, Andrea Donarini, Sonja Koller, and Milena Grifoni }
\affiliation{Institut f{\"u}r Theoretische Physik, Universit{\"a}t Regensburg, D-93040 Regensburg, Germany}
\date{\today}
\begin{abstract}
A theoretical model of a single molecule coupled to many vibronic modes is presented. At low energies, transport is dominated by electron-vibron processes where transfer of an electron through the dot is accompanied by the excitation or emission of quanta (vibrons). Because the frequency of the $n$th mode is taken as an $n$th multiple of the frequency of the fundamental mode, several energetically degenerate or quasidegenerate vibronic configurations can contribute to transport. We investigate the consequences of strong electron-vibron coupling in a fully \emph{symmetric} setup. Several striking features are predicted. In particular, a gate-asymmetry and pronounced negative differential conductance features are observed. We attribute these features to the presence of slow channels originating from the interplay of Franck-Condon suppression of transport channels with spin and/or orbital degeneracies.
\end{abstract}
\maketitle
\section{Introduction}\label{SParticle}
The coupling of electronic and mechanical degrees of freedom is at the core of the physics of nanoelectromechanical systems (NEMS)\cite{ekinci2005}. Pronounced vibrational effects in electronic transport have been observed in several recent experiments on molecules~\cite{park2002,yu2004,qiu2004,park2000,smit2002} and single wall carbon nanotube quantum dots~\cite{sapmaz2006,huetel2009,leturcq09}. Stimulated by the experimental works, several groups ~\cite{braigPRB2003,jkochPRB2006,kochPRL2005,leturcq09,flensbergnlp2006,boseEL2001,izumida2005,CavalierePRB(R)2010} have attempted to theoretically explain some features of the measured stability diagrams (i.e., of the differential conductance in a bias voltage-gate voltage colour map) which appear to be ubiquitous. Specifically, one observes the following: (i) Equidistant lines run parallel to the edges of the Coulomb diamonds. (ii) The probed diamonds show often negative differential conductance (NDC) features appearing between excited vibronic states. (iii) This corresponds to a sequence of peaks in the current as a function of the bias voltage. The height of the peaks measured in Ref.~\onlinecite{sapmaz2006} was found to be in quantitative agreement with predictions of a simple Franck-Condon model for a single electronic level coupled to a harmonic mode (the so called Anderson-Holstein model)~\cite{braigPRB2003,jkochPRB2006,kochPRL2005,mitraPRB2004}. The NDC features have been explained in Ref.~\onlinecite{boseEL2001} in terms of local vibronic excitations. In Ref.~\onlinecite{ShultzPRB2010} NDC features are associated with the Anderson-Holstein model at a single vibronic resonance, while in Ref.~\onlinecite{ShenPRB2007} NDCs appear due to the electron-phonon-coupling-induced selective unidirectional cascades of single-electron transitions.
 In a recent work~\cite{CavalierePRB(R)2010}, carbon nanotube-specific NDC features have been attributed to a spatially dependent Franck-Condon factor and to the presence of quasidegenerate electronic levels, while in Ref.~\onlinecite{schultz2010} an interference effect between the orbitally degenerate electronic states has been proposed.
In Refs.~\onlinecite{CavalierePRB(R)2010},~\onlinecite{ShenPRB2007}, and~\onlinecite{schultz2010}, however, to see NDC some sort of asymmetry is required either in the coupling to
the source and drain contacts or in the coupling to the two orbitally quasidegenerate states of the system or in  both couplings at the same time.\\\indent
In this paper we extend these ideas and propose a generic model in which two degenerate or quasidegenerate molecular levels are coupled to many vibronic modes. We consider the case of a \textit{symmetric} setup, i.e., invariant under exchange of source with drain if simultaneously also the sign of the bias voltage is reversed. If the frequency $\omega_n$ of the $n$th mode is an $n$th multiple, $\omega_n=n\omega$, of the frequency $\omega\equiv\omega_1$ of the fundamental mode, then several energetic degenerate vibronic configurations arise (involving two or more modes) which may contribute to transport at finite bias.\\
As we are going to show in our paper, the additional presence of spin and/or orbital degeneracies opens the possibility of getting slow channels contributing to transport. As a consequence, NDC phenomena can occur despite a \textit{fully symmetric} quantum-dot setup. A peculiarity of the observed features is, in particular, an asymmetry with respect to the gate voltage in the stability diagrams.
Finally, we retain source-drain symmetry but allow for an asymmetry in the coupling to the two orbitally degenerate states. We show that this asymmetry is sufficient to explain the experimental observations (i)-(iii).\\\indent
In some of the experiments~\cite{sapmaz2006,CavalierePRB(R)2010} the slope of the NDC lines is the same for both positive and negative biases. This characteristic has been associated with the left and right asymmetry in the coupling to the leads~\cite{CavalierePRB(R)2010}. We confirm that including the higher harmonics does not change the effect and give an analytical interpretation of the numerical results for low biases.\\\indent
The paper is organized as follows: In Sec. II the model Hamiltonian of a single molecule coupled to several vibronic modes is introduced. A polaron transformation is employed to obtain the spectrum of the system in the presence of electron-vibron interactions. In turn, as known from the theory of Franck-Condon blockade in the simplest Anderson-Holstein model~\cite{braigPRB2003,kochPRL2005}, the polaron transformation also crucially affects the tunneling Hamiltonian describing the coupling to the source and drain leads.\\\indent
In Sec. III the consequences on transport are analyzed. (i) The tunneling transition amplitudes involve product of Franck-Condon factors with coupling constants depending on the mode number. (ii) At low bias, such that only the lowest vibronic mode is excited, a description of the dynamics only in terms of rate equations involving occupation probabilities of the many-body states of the quantum dots is appropriate. (iii) At higher bias, when several vibron modes are excited, a generalized master equation (GME) coupling diagonal (populations) and off-diagonal (coherences) elements of the quantum dot reduced density matrix should be used (see  e.g., Refs.~\onlinecite{schultz2010,begemannPRB2008,donarininanolett2009,braigPRB2005,braunPRB2004,
wunschPRB2005,harbolaPRB2006,mayrhoferEPJB2007,kollerNJP2007,hornberger2008,schultzPRB2009}).\\\indent
In Sec. IV our main results for the current-voltage characteristics in a fully symmetric setup are shown and analyzed. In particular, we give an explanation of the NDC features observed at low bias in terms of a different spin and orbital degeneracy of states with different electron numbers contributing to transport. In Sec. V asymmetric setups are discussed as well. Conclusions are drawn in Sec. VI.
\section{Model Hamiltonian}
In the following we consider a generalized Anderson-Holstein model where the Hamiltonian of the central system has the form
\begin{equation}\label{systemhamiltoian}
\hat{H}_{sys}=\hat{H}_{mol}+\hat{H}_{v}+\hat{H}_{e-v},
\end{equation}
where $\hat{H}_{mol}$ describes two quasidegenerate levels.
The Hamiltonian is modeled as
\begin{equation}\label{modelhamiltoian}
\hat{H}_{mol}=\sum_{l\sigma}\varepsilon_{l}\hat{N}_{l\sigma}+\frac{U}{2}\hat{N}\left(\hat{N}-1\right),
\end{equation}
where  $l=1,2$ is the orbital and $\sigma=\uparrow,\downarrow$ is the spin degree of freedom.
The operator $\hat{N}_{l\sigma}=\hat{d}^\dag_{l\sigma}\hat{d}_{l\sigma}$ counts the number of electrons with spin
$\sigma$ in the orbital $l$. $\hat{N}=\sum_{l\sigma}\hat{N}_{l\sigma}$ is the total number operator. The orbital energy is $\varepsilon_{l}=\varepsilon_0\left[1+\left(-1\right)^l\Delta\right]$ with $\Delta$ an orbital-mismatch. The Coulomb blockade is taken into account via the charging energy $U$ and we assume $U>\varepsilon_0$.\\
The vibron Hamiltonian is expressed as
\begin{align}\label{phononhamiltonian}
 \hat{H}_{v}= \sum_{n\geq 1} \varepsilon_n\left(\hat{a}^\dagger_{n}\hat{a}_{n}+\frac{1}{2}\right),
\end{align}
where $\hat{a}_{n}(\hat{a}^\dagger_{n})$ annihilates (creates) a vibron in the $n\rm{th}$ mode of energy $\varepsilon_n=\hbar\omega_n$. We assume that  the energy of the $n\rm{th}$ mode is given by
\begin{align}\label{phononenergy}
 \varepsilon_n=n\hbar\omega,
\end{align}
being an $n\rm{th}$ multiple of the energy $\varepsilon_1=\hbar\omega$ of the fundamental mode as it is, for example, for longitudinal stretching modes in quantum wires and carbon nanotubes.
Finally, the electron-vibron interaction Hamiltonian is given by
\begin{align}\label{e-phhamiltonian}
 \hat{H}_{e-v}= \sum_{n\geq 1}\sum_{l\sigma} g_n\hat{N}_{l\sigma}\left(\hat{a}^\dagger_{n}+\hat{a}_{n}\right),
\end{align}
where $g_n$ is the coupling constant for the $n\rm{th}$ vibronic mode.
\subsection{Polaron transformation}
 In order to solve the Hamiltonian of the system, we decouple the electron-vibron interaction part by applying a unitary polaron transformation~\cite{mahan2000}.
Specifically, we set $\tilde{\hat{H}}_{sys}\equiv e^{\hat{S}}\hat{H}_{sys}e^{-\hat{S}}$, where
\begin{equation}\label{polaronoperator}
\hat{S}=\sum_{n\geq 1}\sum_{l\sigma} \lambda_n\hat{N}_{l\sigma}\left(\hat{a}^\dagger_{n}-\hat{a}_{n}\right)
\end{equation}
and $\lambda_n=g_n/\hbar\omega_n$ is the dimensionless coupling constant associated with mode $n$. Notice that $\lambda=\frac{g_1}{\hbar\omega_1}$ is the coupling constant of the fundamental mode. We assume $\lambda=0.68,\quad0.83\quad\rm{and}\quad1.18$ in the analysis of the spectrum, which is in the range of values observed  e.g., in experiments on carbon nanotubes~\cite{sapmaz2006,leturcq09,CavalierePRB(R)2010}.
Under the polaron transformation the operator $\hat{d}_{\sigma l}$ is transformed as
\begin{equation}
\tilde{\hat{d}}_{l\sigma}=e^{\hat{S}}\hat{d}_{l\sigma}e^{-\hat{S}}=\hat{d}_{l\sigma}\hat{X},
\end{equation}
where $\hat{X}=\exp\left[-\sum_{n\geq 1}\lambda_n\left(\hat{a}^\dagger_{n}-\hat{a}_{n}\right)\right]$.
In a similar way the shifted vibronic operator is
\begin{align}\label{phononoperator}
\tilde{\hat{a}}_{n}=\hat{a}_{n}- \lambda_n\sum_{l\sigma}\hat{N}_{l\sigma}.
\end{align}
The transformed form of the system Hamiltonian is thus
\begin{align}\label{systemhamiltonian2}
 \tilde{\hat{H}}_{sys}&=\sum_{l\sigma}\tilde{\varepsilon}_l\hat{N}_{l\sigma}+ \sum_{n\geq 1} \varepsilon_n\left(\hat{a}^\dagger_{n}\hat{a}_{n}+\frac{1}{2}\right)\nonumber\\&+\frac{\tilde{U}}{2}\hat{N}\left(\hat{N}-1\right),
\end{align}
where $\tilde{\varepsilon}_l=\varepsilon_l-\sum_n\frac{\left|g_n\right|^2}{\hbar\omega_n}$ is the renormalized orbital energy and $\tilde{U}=U-2\sum_n\frac{\left|g_n\right|^2}{\hbar\omega_n}$ is the Coulomb repulsion modified by the vibron mediated interaction.\\
The eigenstates of the system are
\begin{align}\label{systemeigenstates}
{|\vec{N},\vec{m}_{v}\rangle }_1 := & e^{-\hat{S}}|\vec{N},\vec{m}_{v}\rangle.
\end{align}
where $\vec{N}=\left(N_{1\uparrow},N_{1\downarrow},N_{2\uparrow},N_{2\downarrow}\right)$ and $N_{l\sigma}$ the number
of electrons in the branch $\left(l\sigma\right)$. Notice that $N=\sum_{l\sigma}N_{l\sigma}$ defines the total number of electrons on the dot.
For later purposes we indicate the ground state and first excited state with $0$ electrons as [see Fig.~\ref{fig2}(b)]
\begin{align}\label{4mparticlestates}
&|0,0\rangle:=|\vec{0},\vec{0}\rangle_1,\nonumber\\& |0,1\rangle:=|\vec{0},\vec{m}_{v}=(1,0,0,...)\rangle_1,
\end{align}
The first excited state with $0$-electron contains one vibronic excitation in the first mode, i.e., $\vec{m}_{v}=\left(1,0,0,... \right) $.
In a similar way we define the ground states and first excited states with $N=1$ electron. For zero orbital mismatch one has fourfold degeneracy [see again Fig.~\ref{fig2}(b)], i.e.,
\begin{align}\label{4m+1particlestates}
&|1_k,0\rangle:=|\vec{1}_k,\vec{0}\rangle_1,\quad k=1,2,3,4,  \nonumber\\&|1_k,1\rangle:=|\vec{1}_k,\vec{m}_{v}=(1,0,0,...)\rangle_1,
\end{align}
where $\vec{1}_k \in \bigl\{\left(1,0,0,0 \right), \left(0,1,0,0 \right)$,$\left(0,0,1,0 \right),\left(0,0,0,1 \right) \bigr\}$.
 We notice that both for $N=0$ and $N=1$ the second excited states are vibronically degenerate for the dispersion relation Eq.~\eqref{phononenergy}, then the configurations $\vec{m}_{v}=\left(2,0,0,... \right) $ and $\vec{m}_{v}=\left(0,1,0,... \right) $ have the same energy. For finite orbital mismatch, $\Delta\neq 0$, the orbital degeneracy is broken. The case $0<\varepsilon_0\Delta<\hbar\omega$ is illustrated in Fig.~\ref{fig8}(b).
The corresponding states are:
\begin{align}\label{particlestatesfinitedelta}
&|1_{kg},0\rangle:=|\vec{1}_{kg},\vec{0}\rangle_1,\quad k=1,2,\nonumber\\& |1_{ke},0\rangle:=|\vec{1}_{ke},\vec{0}\rangle_1,\quad k=1,2,\nonumber\\&  |1_{kg},1\rangle:=|\vec{1}_{kg},\vec{m}_{v}=(1,0,0,...)\rangle_1,\nonumber\\&|1_{ke},
1\rangle:=|\vec{1}_{ke},\vec{m}_{v}=(1,0,0,...)\rangle_1,
\end{align}
and $\vec{1}_{kg} \in \left\lbrace{\left(1,0,0,0\right) ,\left(0,1,0,0 \right) }\right\rbrace $,\\  $\vec{1}_{ke} \in \left\lbrace{\left(0,0,1,0 \right) ,\left(0,0,0,1\right) }\right\rbrace$.
\section{Sequential transport}
In this section we discuss the transport across the system under the assumption of weak coupling to the leads. The Hamiltonian of the full system is described as
\begin{align}
{\hat{H}}={\hat{H}}_{sys}+\sum_{\alpha=s,d}\hat{H}_\alpha +{\hat{H}}_T+\hat{H}_{ext},
\end{align}
where $\alpha=s,d$ denote the source and the drain contact, respectively. The tunneling Hamiltonian ${\hat{H}}_T$ is given by
\begin{align}\label{tunnelinghamiltonian}
{\hat{H}}_T=\sum_{\alpha\kappa\sigma}\sum_{ l} \left(t_{\alpha l}\hat{d}^\dag_{l\sigma} \hat{c}_{\alpha\kappa\sigma}+\rm{H.c.} \right),
\end{align}
where $\hat{c}_{\alpha\kappa\sigma}$ is the electron operator in the leads.
Finally, $\hat{H}_{ext}$ describes the influence of the externally applied gate voltage $V_g$. The gate is capacitively coupled to the molecule and hence contributes via a term $eV_g\hat{N} $.
In the case of high degeneracy of the spectrum, the appropriate technique to treat the dynamics of the system in the weak coupling regime is the Liouville equation method for the time evolution of the density matrix of the total system consisting of the leads and the generic quantum dot. To describe the electronic transport through the molecule, we solve the Liouville equation
\begin{align}\label{Liouvilleequation}
i\hbar\frac{\partial\hat{\rho}_{red}^I(t)}{\partial t}=Tr_{leads}\left[{\hat{H}}^I_T(t),\hat{\rho}^I(t) \right],
\end{align}
for the reduced density matrix $\hat{\rho}_{red}(t)=Tr_{leads}\left\lbrace \hat{\rho}(t)\right\rbrace$ in the interaction picture, where the trace over the leads degrees of freedom is taken. We make the following standard approximations to solve the above equation:
(i) The leads are considered as reservoirs of noninteracting electrons in thermal equilibrium. (ii) We factorize the total density matrix as $\rho^I(t)=\rho^I_{sys}(t) \otimes \rho_{leads}$, where we assume weak coupling to the leads, and treat ${\hat{H}}_T$ perturbatively up to second order.
(iii) Being interested in long time properties, we make a Markov approximation, where the time evolution of $\dot{\hat{\rho}}_{red}^I(t)$ is only local in time. Notice that in the regime of interest, $t\longrightarrow\infty$, the Markovian approximation becomes exact.
(iv) Since the eigenstates ${|\vec{N},\vec{m}_{v}\rangle }_1 $ of $ {\hat{H}}_{sys} $ are known, it is convenient to calculate the time evolution of ${\hat{\rho}_{red}}^I $ in this basis retaining coherences between degenerate states with the same number of particles.
Hence  ${\hat{\rho}_{red}}^I $ can be divided into block matrices $\rho^{I,E_N}_{nm}(t)$, where $E$ and $N$ are the energy and number of particles of the degenerate eigenstates $|n\rangle,|m\rangle \in \left\lbrace{|\vec{N},\vec{m}_{v}\rangle }_1 \right\rbrace$.
We obtain an equation of the Bloch-Redfield form\\
\begin{align}\label{LiouvilleequationBloch-Redfieldform}
\dot{\rho}^{I,E_N}_{nm}(t)& =-\sum_{kk'}R^{E_N}_{nmkk'}\rho^{I,E_N}_{kk'}(t)\nonumber\\&+\sum_{M=N\pm1}\sum_{E'}\sum_{kk'}
R^{E_NE'_M}_{nmkk'}\rho^{I,E'_M}_{kk'}(t),
\end{align}
where the indices $n,m,k,k'$ refer to the eigenstates of ${\hat{H}}_{sys}$ and $k,k'$ runs over all degenerate states with fixed particle number. Notice that if $N=0$ then $M=1$, while if $N=4$ then $M=3$. The Redfield tensors are given by~\cite{Blum}
\begin{align}\label{Redfieldtensor}
R^{E_N}_{nmkk'}=&\sum_{\alpha=s,d}\sum_{M,E',j}\left(\delta_{mk'}\Gamma^{(+)E_NE'_M}_{\alpha,njjk}+\delta_{nk}
\Gamma^{(-)E_NE'_M}_{\alpha,k'jjm}\right),
\end{align}
and $R^{E_NE'_M}_{nmkk'}=\sum_{\alpha,p=\pm}\Gamma^{(p)E'_ME_N}_{\alpha,k'mnk}$, where the quantities  $\Gamma^{(p)E_NE'_M}_{\alpha,njjk}$ are transition rates from a state with $N$ to a state with $M$ particles.
More explicitly:
\begin{align}\label{transitionrate1}
\Gamma^{(p)E_NE'_{N+1}}_{\alpha,k'mnk}= \sum_{l\sigma}\Gamma^{\left( p\right) }_{\alpha,l +}\left( \varepsilon_{\alpha}\right)\left(\hat{d}_{l\sigma}\right)^{E_NE'_{N+1}}_{k'm} \left(\hat{d}^\dag_{l\sigma}\right)^{ E'_{N+1}E_N}_{nk},
\end{align}
with $\varepsilon_{\alpha}=-eV_\alpha-\left(E_N-E'_{N+1}\right) $ and $V_\alpha$ the electrochemical potential of the lead $\alpha$. Likewise
\begin{align}\label{transitionrate2}
\Gamma^{(p)E_NE'_{N-1}}_{\alpha,k'mnk}=\sum_{l\sigma}\Gamma^{\left(p\right)}_{\alpha,l-}\left( \varepsilon'_{\alpha}\right)\left(\hat{d}^\dag_{l\sigma}\right)^{E_NE'_{N-1}}_{k'm}\left(\hat{d}_{l\sigma}\right)^{ E'_{N-1}E_N}_{nk},
\end{align}
with $\varepsilon'_{\alpha}=-eV_\alpha-\left(E'_{N-1}-E_{N} \right) $. Moreover, we introduced
\begin{align}\label{transitionrate3}
\Gamma^{\left(p\right)}_{\alpha,l\pm}\left(E\right)=\gamma_{\alpha l}f_\pm\left( E\right) +\frac{ip}{\pi}\gamma_{\alpha l}P\int d\varepsilon \frac{f_\pm\left( \varepsilon\right)}{\varepsilon-E},
\end{align}
where $f_+\left( \varepsilon\right)=f\left( \varepsilon\right) $ is the Fermi function while $f_{-}\left( \varepsilon\right) =1-f\left( \varepsilon\right)$ and $\gamma_{\alpha l}=\frac{2\pi}{\hbar}D_\alpha\left|t_{\alpha l}\right|^2$ are the bare transfer rates with the constant densities of states of the leads $D_\alpha$.
Knowing the stationary density matrix $\rho^I_{st}$, the (particle) current through lead $\alpha$ is determined by $(\alpha=s/d)$
\begin{align}\label{current}
I_\alpha=2\alpha Re\sum_{N,E,E'}\sum_{nkj}\left(\Gamma^{(+)E_NE'_{N+1}}_{\alpha,njjk}-\Gamma^{(+)E_NE'_{N-1}}_{\alpha,njjk}
\right)\rho^{I,E_N}_{kn,st}.
\end{align}
If the relation given by Eq.~\eqref{phononenergy} holds, then spin and orbital degeneracies intrinsic in the electronic structure are supplemented by degeneracies related to the vibronic structure. Several vibronic modes with frequencies $\omega_n = n\omega$ multiples of the fundamental frequency $\omega$ give rise, in fact, naturally to several degenerate vibronic configurations. This is the situation we shall focus on in the rest of the paper.\\\indent
A degenerate spectrum is a necessary condition for the appearance of interference effects in the transport characteristics both in the linear and non linear regime \cite{schultz2010,begemannPRB2008,donarininanolett2009,braigPRB2005,
braunPRB2004,wunschPRB2005,harbolaPRB2006,mayrhoferEPJB2007,kollerNJP2007,
hornberger2008,schultzPRB2009} and these effects can be captured only by considering not only populations (diagonal elements) but also coherences (off-diagonal elements) of the reduced density matrix.\\\indent
For the system at hand we calculated the current both with and without coherences between degenerate states up to five vibronic modes, obtaining though only quantitative but not qualitative differences. While spin and orbital degeneracies can be a priori excluded from the transport through a single molecule with nonpolarized leads\cite{mayrhoferEPJB2007}, the role played by the vibronic coherences requires a more careful analysis.\\\indent
We have confirmed that it is not possible to construct a linear combination of degenerate states $\left\lbrace |s\rangle\right\rbrace $ with finite transition amplitude to a state $|r\rangle$ at one lead  but decoupled from $|r\rangle$ at the other lead, where $|r\rangle$ and $|s\rangle $ represent the states given by Eq.~\eqref{systemeigenstates}. This observation, complemented by the general method presented in Ref.~\onlinecite{donariniPRB2010} proves the absence of interference blocking states in our system. Thus, interference, even if present, does not have dramatic consequences on the transport characteristics of the system.\\\indent
All the current maps presented in the next section are hence, apart from Fig.~\ref{fig6}(b), calculated neglecting coherences. As shown explicitly in Fig.~\ref{fig6}(b), this approximation does not affect qualitatively the results (at least in the low bias regime). Moreover, the negative differential conductance and the associated dynamical symmetry breaking that we present in the next section are not related to the interference and can thus be obtained by considering the dynamics of the populations alone.
\section{Symmetric Setup}
In this section, we illustrate our predictions for the transport characteristics and focus on the $0\leftrightarrow 1$ transitions. In the calculation we also assume for the coupling constant of the $n\rm{th}$ mode $g_n=\sqrt{n}g_1$ (as expected for stretching modes in carbon nanotubes \cite{izumida2005}). The system is \textit{symmetrically} coupled to source and drain contacts ($\gamma_{sl}=\gamma_{dl}$) and the lowest five vibron modes are included. Results for the differential conductance for different values of the dimensionless coupling constant $\lambda$ are illustrated in Fig.~\ref{fig1} and Fig.~\ref{fig7}, obtained for zero and finite orbital mismatch $\varepsilon_{\Delta}=\varepsilon_0\Delta$, respectively.
\subsection{$I-V$ characteristics at low bias and zero band mismatch}\label{Zerobandmismatch}
\begin{figure}[h]
\centerline{\psfig{figure=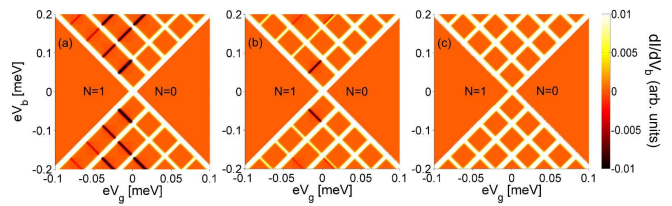,width=\columnwidth}}
\caption{(Color online). (a)-(c) Plots of the numerical differential conductance $dI/dV$(arbitrary units) of the system for coupling constants $\lambda=0.68,\quad 0.83$ and $1.18$, respectively.  The charge excitation energy is $\varepsilon_0=1.4{\rm meV}$ and the energy of the lowest vibron mode is $\varepsilon_1=0.04{\rm meV}$.
Additional parameters are a thermal energy of $k_B T=0.8\mu {\rm eV}$, orbital mismatch $\varepsilon_{\Delta}=0$ and $\gamma_{sl}=\gamma_{dl}=0.02\mu {\rm eV}$ for $l=1,2$. The black lines running parallel to the Coulomb diamond edges correspond to negative differential conductance (NDC). Notice that here and in the following figures $dI/dV$(arbitrary units) is normalized to the maximum of $dI/dV$(arbitrary units) in the considered parameter range. The gate voltage is set to zero by convention at the degeneracy point.}
\label{fig1}
\end{figure}

 When the orbital mismatch is zero, i.e., $\varepsilon_{\Delta}=0$, then the two orbital energies $\varepsilon_l$ are the same. The minimum energy to produce a charge excitation is $\varepsilon_0$. We take the value of this energy as $\varepsilon_0=1.4 {\mathrm meV}$ (comparable to the level spacing energy of a suspended single wall carbon nanotube of $1.2\mu m$ length). Furthermore, we assume the energy of the lowest vibronic mode to be $\varepsilon_1=0.04{\rm meV}$. Thus the charge excitation energy is much larger than the energy of the lowest vibron mode. Indeed all the equidistant lines running parallel to the diamond edges observed in Fig.~\ref{fig1} are due to vibron excited states. What striking is the occurrence of negative differential conductance (NDC) features at moderate coupling ($\lambda=0.68$ and $\lambda=0.83$) which, however, disappear when the coupling is increased ($\lambda=1.18$). Moreover, the NDC lines are only running parallel to one of the diamond edges, which indicates an asymmetry with respect to the gate voltage $V_g$. As we are going to explain, at low bias, these features are a consequence of Franck-Condon assisted tunneling combined with the spin and/or orbital degeneracy in the system.

\begin{figure}[h]
\centerline{\psfig{figure=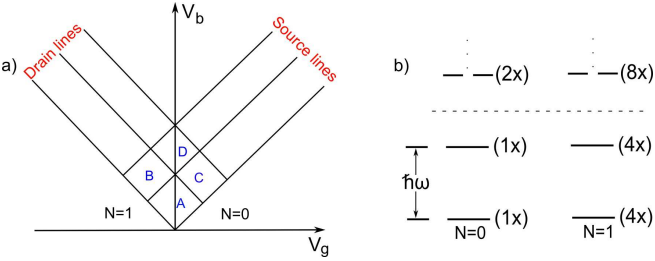,width=\columnwidth}}
\caption{(Color online).  (a) The low-bias transition regions of the stability diagram are labelled as A, B, C, D. (b) Energy level scheme for the relevant transitions in the stability diagram involving regions A-D. Above the dashed line region D is activated with two vibron modes being in the transport window. The energy of the lowest vibron mode is $\hbar\omega = 0.04{\rm meV}$. The number of degenerate states is indicated in bracket.}
\label{fig2}
\end{figure}

\begin{figure}[h]
\centerline{\psfig{figure=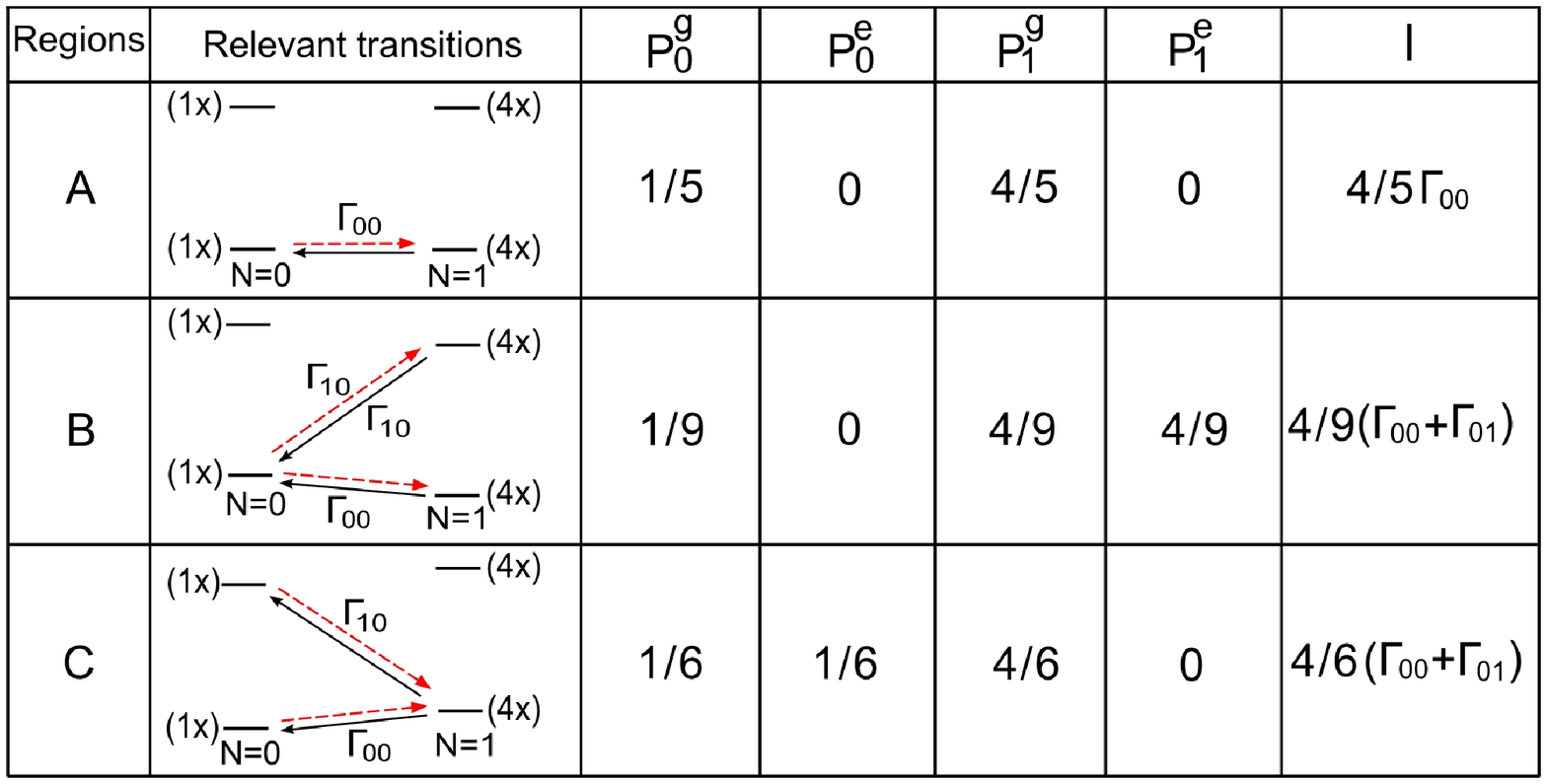,width=\columnwidth}}
\caption{(Color online).  In the table for each of the different regions A, B, C in the stability diagram, relevant transitions, population of states and current are given.  $P^g_{0}$, $P^e_{0}$ represent the population of the $0$-particle ground and  first excited states, respectively, and $P^g_{1}$, $P^e_{1}$ the population of the $1$-particle ground and first excited states.  $I$ is the corresponding current in  each region. In the transition scheme, the black arrows represent the drain and the dashed red arrows the source transitions. $\Gamma_{00}$ denotes the transition rate from $0$-particle ground state to the $1$-particle ground state while $\Gamma_{01}$ the transition rate from the $0$-particle ground state to a $1$-particle first excited state.}
\label{fig3}
\end{figure}

\begin{figure}[h]
\centerline{\psfig{figure=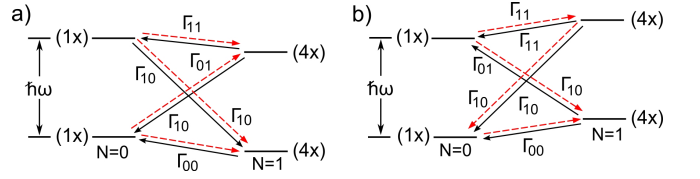,width=\columnwidth}}
\caption{(Color online).  (a) Energy level scheme for transitions in region B and (b) in region C. \hlS{Importantly, because the bias voltage is too low, the transition \hlA{$|1_k,0\rangle\to|0,1\rangle$} in region B and the transition $|0,0\rangle\to|1_k,1\rangle$ in region C are not allowed.}}
\label{fig4}
\end{figure}

\begin{figure}[h]
\centerline{\psfig{figure=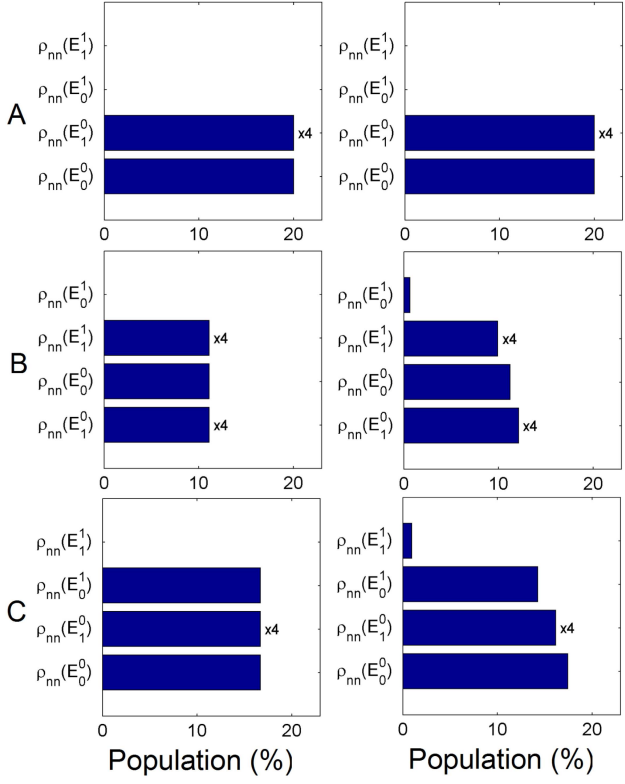,width=9.5cm}}
\caption{(Color online).  Populations of the low energy states corresponding to the stationary density matrix calculated for different electron vibron coupling $\lambda$ and different gate-bias ranges. The first column corresponds to the case $\lambda=1$, thus $\Gamma_{11} = 0$, while in the second column $\lambda = 0.83$. The letters A, B, C labeling the rows refer to the stability diagram regions defined in Fig.~\ref{fig2}. The states are ordered in energy. The rest of the parameters are the same as used for Fig.~\ref{fig1}.}
\label{fig5}
\end{figure}

\begin{figure}[h]
\centerline{\psfig{figure=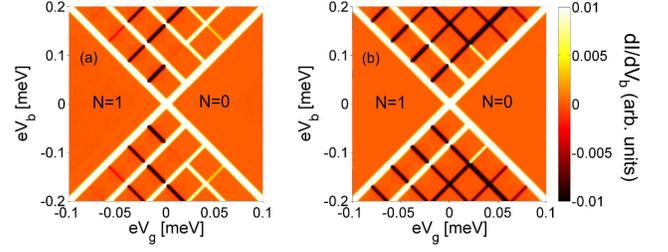,width=\columnwidth}}
\caption{\hlA{(Color online). Plots of the differential conductance for a coupling constant $\lambda=0.68$ with two different approximations: (a) neglecting the higher harmonics of the system vibrations, (b) keeping coherences between the degenerate states with different vibronic configurations. The rest of the parameters are the same as used for Fig.~\ref{fig1}(a).}}
\label{fig6}
\end{figure}

Specifically, let us focus on the low bias region [see Fig.~\ref{fig2}(a)], where only ground-state $\leftrightarrow$ ground-state transitions (region A), and ground-state $\leftrightarrow$ first excited-state transitions (regions B, C) are relevant. The $0$ and $1$-particle states involved are illustrated in Fig.~\ref{fig2}(b), together with their degeneracy due to spin and orbital degrees of freedom, and have energies below the dashed line in Fig.~\ref{fig2}(b). The states above the dashed line require an energy of at least $2\hbar\omega$ and have thus also a vibron degeneracy.
In the considered energy range no degenerate vibron configurations are involved. Moreover, coherences between degenerate electronic configurations are not present such that a rate equation description only in terms of populations is appropriate. At low bias and in the stationary limit Eq.~\eqref{LiouvilleequationBloch-Redfieldform} yields the equation for the populations:
\begin{align}\label{populations}
0=\dot{\rho}^{E_N}_{nn} =-R^{E_N}_{nnnn}\rho^{E_N}_{nn}+\sum_{M=N\pm 1}\sum_{E'}\sum_{k}R^{E_NE'_M}_{nnkk}\rho^{E'_M}_{kk},
\end{align}
or, equivalently,
\begin{align}\label{populations2}
0=&\sum_{M=N\pm 1}\sum_{E'}\sum_{k}\sum_\alpha\left(\Gamma^{E_N E'_M}_{\alpha,nk}\rho^{E_N}_{nn}-\Gamma^{E'_ME_N}_{\alpha,kn}\rho^{E'_M}_{kk}\right),
\end{align}
where $\Gamma^{E_N E'_M}_{\alpha,nk}\equiv 2Re\Gamma^{(+)E_NE'_M}_{\alpha,nkkn}$ and $\Gamma^{E'_ME_N}_{\alpha,kn}\equiv 2Re\Gamma^{(+)E'_ME_N}_{\alpha,knnk}$, see Eqs. (\ref{transitionrate1})-(\ref{transitionrate3}).
Notice in particular that $\Gamma^{E_N E'_{N+1}}_{\alpha,nk}=\gamma_{\alpha l} f_+\left( eV_\alpha-\left(E'_{N+1}-E_N\right) \right)C_{nk}$ and $\Gamma^{E'_{N+1} E_N}_{\alpha,kn}=\gamma_{\alpha l}f_-\left( eV_\alpha-\left( E'_{N+1}-E_N\right) \right)C_{nk} $, i.e., they only differ in the Fermi factors.
The transition coefficients are the same and given by $C_{nk}=\sum_{l\sigma}\left|\left( \hat{d}_{l\sigma}\right)^{E_N E'_{N+1}}_{kn} \right|^2$.
Moreover, we find for the current through lead $\alpha$
\begin{align}\label{current2}
I_\alpha=&\alpha\sum_{N,E,E'}\sum_k\sum_{n}\left( \Gamma^{E_N E'_{N+1}}_{\alpha,nk}-\Gamma^{E'_{N+1} E_N}_{\alpha,kn}\right)\rho^{E_N}_{nn}.
\end{align}
Let us now focus on region A. In this case only the $0$-particle ground state $|0,0\rangle$ [see Eq.~\eqref{4mparticlestates}] with energy $E^0_{0}$ and the four $1$-particle ground states $|1_k,0\rangle$ [see Eq.~\eqref{4m+1particlestates}] with energy $E^0_{1}$ contribute to transport. Moreover, inside region A it holds
\begin{align}\label{Fermifunction}
&f\left(eV_s-\left(E^0_{1}-E^0_{0}\right) \right) =1,\nonumber\\& 1-f\left(eV_d-\left(E^0_{1}-E^0_{0}\right) \right) =1,
\end{align}
such that, if $\gamma_{\alpha 1}=\gamma_{\alpha 2}=\gamma_\alpha$, $\gamma_{s}=\gamma_{d}$, it also follows $\left( |n\rangle=|0,0\rangle,\quad |k\rangle\in \left\lbrace{|1_k,0\rangle } \right\rbrace\right) $
\begin{align}\label{groundstatestransitions}
\sum_\alpha \Gamma^{E^0_{0}E^0_{1}}_{\alpha,nk}=\sum_\alpha \Gamma^{E^0_{1}E^0_{0}}_{\alpha,kn}\equiv\Gamma_{00}.
\end{align}
This situation is illustrated in the table of Fig.~\ref{fig3}, where a dashed red (black) arrow indicates a transition  involving the source (drain). Condition (\ref{groundstatestransitions}) with (\ref{populations2}) then implies that
\begin{align}\label{populations3}
\rho^{E^0_{0}}_{nn}=\rho^{E^0_{1}}_{kk} \quad \forall k,\,\text{and}\,\, n
\end{align}
and hence $P^g_{0}:=\rho^{E^0_{0}}_{nn}=\frac{1}{5}$; $P^g_{1}=\sum_k\rho^{E^0_{1}}_{kk}=\frac{4}{5}$,
yielding with Eq.~\eqref{current2} for the current in region A, $I_A=\frac{4}{5}\Gamma_{00}$. Along similar lines we can calculate the current in regions B and C. Let us start with region B where (see table of Fig.~\ref{fig3}) the gate voltage $V_g$ is such that the $1$-particle ground states $|1_k,0\rangle $ have energy $E^0_{1}$ smaller than the one, $E^0_{0}$, of the $0$-particle ground state $|0,0\rangle$. Moreover, in this region also the first excited state $|1_k,1\rangle $ with energy $E^1_{1}$ enters the transport window. We also assume that the rate $\Gamma_{11}$ between the states $|0,1\rangle $ and $|1_k,1\rangle $ is negligible with respect to $\Gamma_{00}$ and $\Gamma_{01}$. Corrections due to a finite $\Gamma_{11}$ will be discussed later. Inside region B it holds, besides Eq.~\eqref{Fermifunction}, and hence Eq.~\eqref{populations3}, $f\left(eV_s-\left(E^1_{1}-E^0_{0}\right)  \right)=1$ and $1-f\left(eV_d-\left(E^1_{1}-E^0_{0}\right)  \right)=1$. Hence it follows that $\left( |n\rangle=|0,0\rangle,\quad |k\rangle \in \bigl\{|1_k,1\rangle\bigr\}\right) $
\begin{align}\label{groundtoexcitedstatestransitions}
\Gamma_{01}\equiv\sum_\alpha \Gamma^{E^0_{0}E^1_{1}}_{\alpha,nk}=\sum_\alpha \Gamma^{E^1_{1}E^0_{0}}_{\alpha,kn}\equiv\Gamma_{10}.
\end{align}
Eq.~\eqref{populations2} implies thus that in region B it holds
\begin{align}
\rho^{E^0_{0}}_{nn}=\rho^{E^0_{1}}_{kk}=\rho^{E^1_{0}}_{kk}=\frac{1}{9},
\end{align}
and hence $P^g_{0}\equiv\rho^{E^0_{0}}_{nn}=\frac{1}{9}$; $P^g_{1}=\sum_k\rho^{E^0_{1}}_{kk}=\frac{4}{9}$; $P^e_{1}=\sum_k\rho^{E^1_{1}}_{kk}=\frac{4}{9}$.
The total current in region B follows from Eq.~\eqref{current2} and reads $I_B=\frac{4}{9}\left(\Gamma_{00}+\Gamma_{01} \right) $. The condition to observe NDC is that $I_B<I_A$, which implies
\begin{align}\label{NDCcondition1}
\Gamma_{01}<\frac{4}{5}\Gamma_{00}.
\end{align}
Along similar lines (see table in Fig.~\ref{fig3}), one finds for the transition from region A to C that $I_C<I_A$ if
\begin{align}\label{NDCcondition2}
\Gamma_{01}<\frac{1}{5}\Gamma_{00}.
\end{align}
Let us look in more detail at Eq.~\eqref{NDCcondition1} and Eq.~\eqref{NDCcondition2}. The rates $\Gamma_{00}$ and $\Gamma_{01}$ describe transitions between states which only differ in their vibronic part. From Eqs. (\ref{groundstatestransitions}), (\ref{groundtoexcitedstatestransitions}) and (\ref{transitionmatrixappendix1}) it follows that
\begin{align}\label{transitionratios}
\frac{\Gamma_{01}}{\Gamma_{00}}=F^2\left( \lambda,0,1\right)=\lambda^2.
\end{align}
Hence, to observe NDC for the transition from region A to B one needs that $\lambda^2<\frac{4}{5}$. On the other hand for NDC in the transition from A to C we must require $\lambda^2<\frac{1}{5}$. Indeed, as shown in Fig.~\ref{fig1}, NDC for the transition A $\leftrightarrow$ B is observed for $\lambda=0.68$ and $\lambda=0.83$, but it vanishes for $\lambda=1.18$. On the other hand NDC is never observed for the transition region A $\leftrightarrow $ C.
Let us now turn to region B and to a finite  $\Gamma_{11}\equiv \sum_\alpha\Gamma^{E^1_{0}E^1_{1}}_{\alpha,nk}=\sum_\alpha\Gamma^{E^1_{1}E^1_{0}}_{\alpha,kn}$ with $|n\rangle =|0,1\rangle $, $k\in \left\lbrace |1_k,1\rangle \right\rbrace$.
Because now $|0,1\rangle $ can get populated, also transitions from $|0,1\rangle$ to $|1_k,0\rangle$ are activated (see Fig.~\ref{fig4}).\\
Because of \hlS{$E^1_{0}-E^0_{1}>eV_s,eV_d$} it holds
\begin{align*}
\tilde{\Gamma}_{10}\equiv \sum_\alpha\Gamma^{E^1_{0}E^0_{1}}_{\alpha,nk}\neq \tilde{\Gamma}_{01}\equiv \sum_\alpha\Gamma^{E^0_{1}E^1_{0}}_{\alpha,kn}=0.
\end{align*}
Hence, the stationary solution with equal probabilities is spoiled, at finite $\Gamma_{11}$, due to the inequality of $\tilde{\Gamma}_{01}$ and $\tilde{\Gamma}_{10}$. In fact, in the case $\Gamma_{11} = 0$, the same inequality only implies that $\rho^{E^1_{0}} = 0$. We also notice that $\tilde{\Gamma}_{10}=\Gamma_{10}$. Moreover, the rates $\Gamma_{11}$ and $\Gamma_{00}$ only differ in their vibronic configuration:
it holds [cf. Eq.~\eqref{bosoncoupling}]
\begin{align*}
\frac{\Gamma_{11}}{\Gamma_{00}}=\left[\sum^1_{i=0}\left(-\lambda^2 \right)^i  \right]^2=\left( 1-\lambda^2\right)^2.
\end{align*}
Likewise
\begin{align*}
\frac{\Gamma_{11}}{\Gamma_{01}}=\frac{\left( 1-\lambda^2\right)^2}{\lambda^2}.
\end{align*}
Hence, if $|\lambda|\approx 1$ it is indeed $\Gamma_{11} \ll \Gamma_{00}, \Gamma_{01}$ and an expansion to lowest order in the ratios $\frac{\Gamma_{11}}{\Gamma_{00}},\frac{\Gamma_{11}}{\Gamma_{01}}$ can be performed.
In this case the conditions for NDC acquire a more complicated form. The condition to get NDC in the source threshold lines [from A to B in Fig.~\ref{fig2}(a)] is
\begin{equation}
\Gamma_{01}< \frac{4}{5}\Gamma_{00}-\frac{23}{40}\Gamma_{11},
\label{eq:AB_nodelta}
\end{equation}
while the condition for NDC in the drain threshold lines [from A to C in Fig.~\ref{fig2}(a)] is
\begin{equation}
\Gamma_{01}< \frac{1}{5}\Gamma_{00}-\frac{7}{10}\Gamma_{11}.
\label{eq:AC_nodelta}
\end{equation}
It means that the presence of chain transition processes redistributes the population among the many-body states in a way that privileges the low energy states (see Fig.~\ref{fig5}); this in turn weakens NDC since it privileges the conducting channels that carry more current.
\hlA{Eventually, let us consider explicitly the effects of the higher harmonics and of the coherences between states with different vibronic configuration on the transport characteristics of the system. In Fig.~\ref{fig6}(a) we present the stability diagram for a coupling constant $\lambda=0.68$ in which we artificially neglect the higher harmonics. By a direct comparison with figure Fig.~\ref{fig1}(a), it is clear that this approximation only marginally affects the NDC and positive differential conductance (PDC) pattern, thus confirming the dominant role played by the spin and pseudospin (orbital) degeneracies in the gate asymmetry. The effect of the coherences, shown in Fig.\ref{fig6}(b), is more complex. Nothing changes for the lowest transition lines where no degeneracy is involved. For higher biases, though, some drain transition lines change their character from PDC to NDC. Thus, the gate asymmetry introduced by the spin and orbital degeneracy and the corresponding NDC (PDC) character of the source (drain) transition lines is exact in the low bias limit but should be taken only as a trend when several excited vibronic states participate in the transport.}
\subsection{$I-V$ characteristics at low bias and finite band mismatch}
\label{sec:LowBias_FinMis}
\begin{figure}[h]
\centerline{\psfig{figure=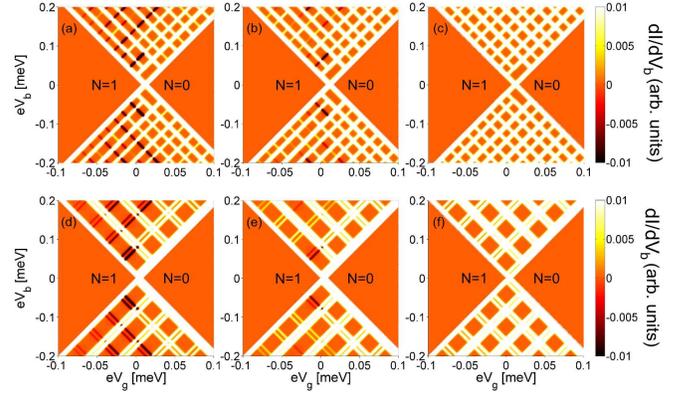,width=\columnwidth}}
\caption{(Color online).  (a)-(c) Stability diagrams for coupling constants $\lambda=0.68,\quad 0.83$ and $1.18$, respectively. Additional parameters are a thermal energy of $k_{\rm{B}} T=0.8\mu {\rm eV}$, orbital mismatch $\varepsilon_{\Delta}=0.016\varepsilon_0$\hlS{$=0.56\varepsilon_1$} and $\gamma_s=\gamma_d=0.02\mu {\rm eV}$ while for (d)-(f) $\varepsilon_{\Delta}=0.006\varepsilon_0$. The rest of the parameters are the same as used for Fig.~\ref{fig1}.}
\label{fig7}
\end{figure}

\begin{figure}[h]
\centerline{\psfig{figure=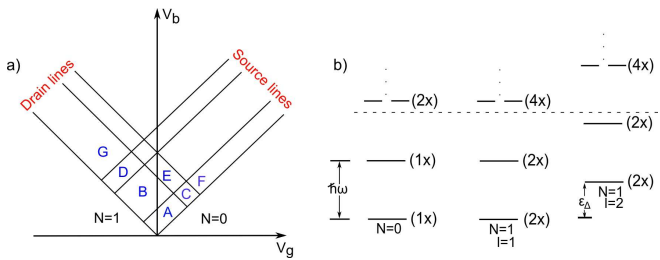,width=\columnwidth}}
\caption{(Color online).  a) The low-bias transition regions are labelled as A, B, C, D, E, F, G. b) Energy level scheme for the transitions relevant in the low bias regions of the stability diagram. As in Fig.~\ref{fig2} is $\hbar\omega = 0.04{\rm meV}$. The degeneracy of each state has been shown in brackets.}
\label{fig8}
\end{figure}

\begin{figure}[h]
\centerline{\psfig{figure=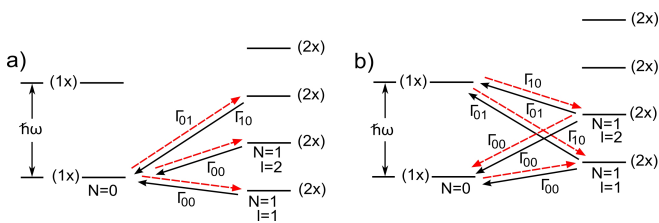,width=\columnwidth}}
\caption{(Color online).  (a) Transition scheme for region D at $\varepsilon_{\Delta}=0.016\varepsilon_0$.  (b) Transition scheme for region F at $\varepsilon_{\Delta}=0.016\varepsilon_0$.}
\label{fig9}
\end{figure}

In this section, we discuss our results on vibration-assisted transport with the same parameters as in Sec.~\ref{Zerobandmismatch} but with a finite orbital mismatch, i.e, $\varepsilon_{\Delta}\neq 0$. In this case the orbital degeneracy is broken. The corresponding stability diagrams are shown in Fig.~\ref{fig7}. The analysis for the NDC conditions at low bias remains almost the same as before. Slight differences occur because in this case the orbital degeneracy is lost and the populations are redistributed over the many-body states in a different way.
In Fig.~\ref{fig8}(a), the different transition regions of the stability diagram have been labeled while the energy level scheme has been shown in Fig.~\ref{fig8}(b), where the degeneracy of each state is given in brackets. We again truncate the process at the dashed line to analyze the lowest-energy excitations. As before $l=1$ and $l=2$ are the orbital degrees of freedom. The transition scheme for regions D and F is shown explicitly in Fig.~\ref{fig9}.
The current-voltage characteristics \hlS{Fig.~\ref{fig7}(a)-(c) and~\ref{fig7}(d)-(f)} are qualitatively the same as far as the mismatch is in the moderate regime $k_{\rm B}T\ll\varepsilon_{\Delta} <\hbar \omega$, the only difference being the position of the resonance lines, that depends on the specific position of the energy levels. In other terms, despite the size and the position of the regions of the stability diagram (Fig.~\ref{fig7}) depend on the mismatch $\varepsilon_{\Delta}$, the value of the current in each region is independent of it. Thus a unified treatment of the two cases presented in Fig.~\ref{fig7} is allowed, despite their apparent qualitative differences.
In particular, we can observe that the current in region B is larger than the current in region A since a new transport channel is opening with the same geometrical coupling ($\Gamma_{00}$) when passing from A to B. This implies that the first source threshold transition line is always a positive differential conductance (PDC) line.
The current in C is equal to the one in A since, due to energy conservation, no new transport channel is opening when passing from A to C. The corresponding resonance line is thus invisible in the stability diagram (see Fig.~\ref{fig7}). The transition that defines the threshold line separating A from C ($|1_{ke},0\rangle \leftrightarrow |0,1\rangle$ at the drain) involves, in fact, states  which are not populated in that bias and gate voltage range.
Finally, the comparison between currents in the adjacent B and D regions and between the currents in the C and F regions results in conditions for the appearance of NDC lines which are very similar to the one in absence of mismatch. In particular, the condition for NDC at the transition between regions B and D is identical to the one for the transition between regions A and the B corresponding to zero mismatch given in Eq.~\eqref{eq:AB_nodelta}. The NDC condition for the transition between region C and F reads instead
\begin{equation}
\Gamma_{01} < \frac{\sqrt{57}-1}{28}\Gamma_{00} + \frac{1}{4}\left(1-\frac{3}{\sqrt{57}}\right)\Gamma_{11},
\label{eq:CF_delta}
\end{equation}
to be compared with the one for the transition between the regions A and C and zero mismatch given in Eq.~\eqref{eq:AC_nodelta}. A similar analysis can be repeated for higher-energy transitions which participate in the transport for higher biases.
It is already clear though from the low energy transitions that a moderate breaking of the orbital degeneracy introduced by the finite mismatch $\varepsilon_{\Delta}$ does not change qualitatively the transport characteristics of the system. In particular, it preserves the presence or absence of asymmetric NDC lines as a function of the electron vibron coupling $\lambda$ (compare Figs.~\ref{fig1} and~\ref{fig7}).
\subsection{Effect of an asymmetric coupling of the different orbital states to the leads}
\label{sec:Asymmetry_PS}
\begin{figure}[h]
\centerline{\psfig{figure=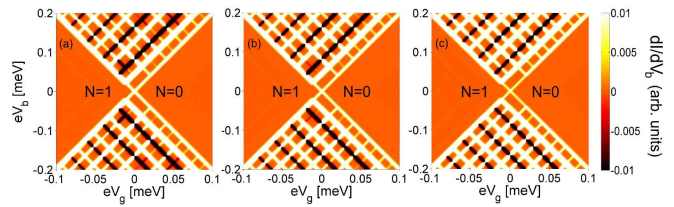,width=\columnwidth}}
\caption{(Color online).  (a)-(c) Stability diagrams for a molecule for the case of a coupling to the leads which depends on the orbital degree of freedom. All the parameters are the same as used in Figs.~\ref{fig1} and~\ref{fig7}. The asymmetry is $a=1/45$ with orbital mismatch $\varepsilon_{\Delta}=0.016\varepsilon_0$.}
\label{fig10}
\end{figure}

\begin{figure}[h]
\centerline{\psfig{figure=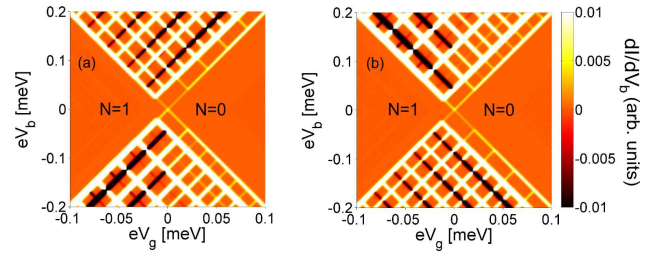,width=\columnwidth}}
\caption{\hlA{(Color online).  (a),(b) Stability diagrams for both an orbital and left and right asymmetry. All the parameters are the same as used in Fig.~\ref{fig10}(c). The asymmetry with respect to the left and right lead is $b=45$ in panel (a) while $b=1/45$ in panel (b).}}
\label{fig11}
\end{figure}

 The Franck Condon factors are still assumed to be the same for the source and drain tunneling and no overall asymmetry is introduced in the tunneling coupling of the molecule dot to the source and drain ($\gamma_{s,1/2} = \gamma_{d,1/2}$). The theory can produce, though, alternating PDC and NDC traces as discussed in Ref.~\onlinecite{CavalierePRB(R)2010} if we assume that coupling of the $l=1$ and $l=2$ orbitals to be different.
 In Fig.~\ref{fig10} we have plotted the differential conductance $\left( dI/dV\right)$ for an asymmetry parameter $a=\frac{\gamma_{\alpha,1}}{\gamma_{\alpha,2}} = 1/45$ where ``1'' , ``2'' represent the orbital degrees of freedom, respectively and $\alpha$ means source or drain. For convenience, in the numerical calculations, we use parameters as in Figs.~\ref{fig1} and~\ref{fig7}.
As seen by comparing Fig.~\ref{fig7} with Fig.~\ref{fig10}, at $\lambda = 1.18$ NDC can now occur. Moreover, an alternation of PDC with NDC lines, as seen in the experiments~\cite{sapmaz2006,leturcq09,CavalierePRB(R)2010} occurs. Repeating the same analysis as in Sec.~\ref{sec:LowBias_FinMis}, we indeed find that: (i) the transition from region A to B gives a NDC line for \hlA{$a > 3/2$} independent of the value of $\lambda$. (ii) the condition governing the transition from region B to D is now modified to be (at $\Gamma_{11} = 0$)
\begin{equation}
\frac{\Gamma_{01}^{(2)}}{\Gamma_{00}^{(2)}} = \lambda^2 <\frac{2+2a}{5a},
\end{equation}
which, for $a<1$, increases the range of $\lambda$ giving NDC also for $\lambda = 1.18$ [Fig.~\ref{fig10}(c)]. ${\Gamma_{00}^{(l)}} $ and ${\Gamma_{01}^{(l)}} $ are defined analogously ${\Gamma_{00}} $ and ${\Gamma_{01}} $ by considering the $l$ dependence of the bare tunneling rates $\gamma_l$. (iii) the transition from region D to G is governed by the condition:
\begin{equation}
 \frac{\Gamma_{01}^{(2)}}{\Gamma_{00}^{(2)}} = \lambda^2 < \frac{2(a+1)}{7-2a},
\end{equation}
which explains the persistence of a PDC line also for smaller values $\lambda$ (compare again Fig.~\ref{fig7} with Fig.~\ref{fig10}). The corrections introduced by a finite $\Gamma_{11}$ rate do not change qualitatively the analysis and can be found for completeness in the Appendix \ref{app:Finite_G11}.
\hlA{
\section{Effect of the asymmetric coupling to the left and right lead}
\label{sec:Asymmetry_LR}
Eventually, we consider the effect of an asymmetry in the coupling to the left and right lead in combination to the orbital asymmetry discussed in the previous section. We introduce the asymmetry via the parameter $b = \gamma_{L,l}/\gamma_{R,l}$ where $l=1,2$ represents the orbital degree of freedom and by convention $\gamma_{s,l} \equiv \gamma_{L,l}$ for positive bias voltages. In Fig.~\ref{fig11} we present the stability diagrams for a molecule coupled to vibrons with an orbital asymmetry $a = 1/45$ and two different left and right asymmetries. In Fig.~\ref{fig11}(a) the asymmetry parameter $b = 45$ while $b = 1/45$ in Fig.~\ref{fig11}(b). Since $b$ is the only parameter of the system that breaks the left and right symmetry, the differential conductances in Fig.~\ref{fig11} can be obtained one from the other by a reflection of the bias. The most striking effect of the left and right asymmetry is, though, to produce the NDC lines always in the same direction  [compare Figs.~\ref{fig10}(c) and~\ref{fig11}(a)] both for positive and negative biases.
As in the previous sections, we studied analytically the transitions lines separating the A, B, D and G low-bias regions of the stability diagram (see Fig.~\ref{fig8}). We could thus obtain the NDC conditions for arbitrary values of the asymmetry parameters $a$ and $b$. The transition line  between region A and B is an
NDC line for every electron phonon coupling $\lambda$ under the condition
\begin{equation}
a > \frac{2b + 1}{2b}.
\label{eq:2asymAB}
\end{equation}
The NDC condition for the transition between the B and D region reads instead
\begin{equation}
\lambda^2 < \frac{2(a+1)}{a}\frac{b}{1+4b},
\label{eq:2asymBD}
\end{equation}
while the transition between the regions D and G is governed by the relation
\begin{equation}
\lambda^2 < \frac{2(a+1)b}{1+2(3-a)b}.
\label{eq:2asymDG}
\end{equation}
Equations \eqref{eq:2asymAB}-\eqref{eq:2asymDG} allow a partial interpretation of the numerical results presented in Fig.~\ref{fig11}. It is in fact easy to demonstrate that, if $b > 1$, for sufficiently small values of $a$ ($ a \lesssim  1/b$) the alternating NDC and PDC pattern at positive biases is not modified by the left and right asymmetry parameter $b$ . With the help of the same set of equations and the symmetry property mentioned above we can also analyze the negative bias transitions. The sequence of transitions between the regions A, B, D, and G in Fig.~\ref{fig11}(b) reveals in fact a very different pattern of strong PDC transitions alternated by very weak NDC lines. This sequence can be obtained by the conditions expressed in Eqs. \eqref{eq:2asymAB}-\eqref{eq:2asymDG} by substituting $b$ with $1/b$ in the limit $b \gg 1$.
Unfortunately, due to the peculiar structure of the harmonic spectrum, the number of states involved in the regions E and F of the  stability diagram grows rapidly and the analytical analysis of the transition, even if possible, becomes very cumbersome. Our numerical findings are nevertheless consistent with the ones reported by other groups \cite{CavalierePRB(R)2010}, where the left and right asymmetry has been given as a necessary condition for achieving NDC with the same slope at both positive and negative biases.}
\section{Conclusions}
In this paper we analyzed the spectrum and the vibration-assisted transport properties of a nanostructure where two degenerate or quasidegenerate levels are coupled to several vibronic modes. Our model can capture features of transport properties of suspended carbon nanotubes and of molecules with a fourfold degenerate electronic level coupled to many vibrational modes. The transport theory is based on vibron-assisted tunneling, mediated by vibrational modes. 
In order to study the dynamics of the system, we apply a density matrix approach which starts from the Liouville equation for the total density operator and which enables the treatment of degenerate and quasidegenerate states.
Despite the fact that we considered a fully symmetric setup, the stability diagram for the differential conductance shows striking negative differential conductance (NDC) features which hints at peculiar features of our nanoelectromechanical system.
We predict that NDCs appear due to the slow channels in the source transition originating from spin and/or orbital degeneracies and the suppression of Franck-Condon channels. With source-drain symmetry being preserved but an asymmetry between orbitally degenerate states being allowed, we could explain the alternating PDC and NDC features observed in Refs.~\onlinecite{sapmaz2006} and~\onlinecite{leturcq09}.
\hlA{Eventually, with the further introduction of the left and right asymmetry suggested in Ref.~\onlinecite{CavalierePRB(R)2010}, we confirmed the appearance, also in presence of multimodes, of the NDC lines with the same slope for both positive and negative biases. We also gave an analytical interpretation of the numerical results in the low-bias regime.}
\section*{Acknowledgments}
We acknowledge the support of DFG under the programs SFB 689 and GRK 1570. A. Y. acknowledges the support of Kohat University of Science \& Technology, Kohat-26000, Khyber Pakhtunkhwa, Pakistan under the Human Resource Development Program. We thank Georg Begemann and Leonhard Mayrhofer for their help in the numerical calculations.
\appendix
\section{Evaluation of transition matrix elements of electron operator}\label{transitionrates}
To determine the transition rates, we calculate the matrix elements
\begin{align}\label{transitionmatrixappendix1}
&\langle r\arrowvert \hat{d}_{l\sigma}\arrowvert s\rangle  =e^{-\frac{1}{2}\sum_n\left|\lambda_n\right|^2}\prod_nF\left(\lambda_{n},m_{n},m'_{n}\right),
\end{align}
where $\arrowvert r\rangle $ and $\arrowvert s\rangle $ represent the eigenstates given by Eq.~\eqref{systemeigenstates}. The function $F(\lambda,m,m')$ determines the coupling between states with a different vibronic number of excitations with effective coupling $\lambda$ and is expressed as
\begin{align}\label{bosoncoupling}
 & F(\lambda, m, m')= \left(\Theta(m'-m){\lambda}^{m'-m}+\Theta(m-m')\right.\nonumber\\&\left.\times{(-\lambda^\ast)}^{m-m'}\right)\sqrt{\frac{{m_{min}}!}{{m_{max}}!}}\sum^{m_{min}}_{i=0}\frac{\left({-|\lambda|^2} \right)^{i} }{i!(i+m_{max}-m_{min})!} \nonumber\\&\times\frac{m_{max}!} {(m_{min}-i)!},
\end{align}
where $m_{min/max}=min/max(m,m')$.

\section{NDC and/or PDC threshold conditions with finite $\Gamma_{11}$}\label{app:Finite_G11}
In this appendix we give the conditions which determines the sign of the current change in the transition from region B to D and D to G with finite mismatch $\varepsilon_{\Delta}$ (Fig.~\ref{fig8}). We take only the first order contribution in the ratios $\Gamma_{11}/\Gamma_{00}$ and $\Gamma_{11}/\Gamma_{01}$. The validity of these formulas is thus restricted to $\lambda \approx 1$. The condition for the transition B to D reads
\begin{equation}
\Gamma_{01}^{(2)}<\frac{2+2a}{5a}\Gamma^{(2)}_{00}-\frac{14+9a}{20(1+a)}\Gamma^{(2)}_{11},
\end{equation}
while for the transition D to G one obtains
\begin{equation}
\Gamma_{01}^{(2)}  < \frac{2(a+1)}{7-2a}\Gamma^{(2)}_{00} -
\frac{4a^2 - 16a - 11}{4(a+1)(2a-7)}\Gamma^{(2)}_{11},
\end{equation}
 where $a = \gamma_{\alpha 1}/\gamma_{\alpha 2}$ measures in both cases the asymmetry between the coupling  to the different orbitals.

\end{document}